\documentclass[12pt]{article}
\usepackage[utf8]{inputenc}
\usepackage[T1]{fontenc}
\usepackage{amsmath}
\usepackage{graphicx}
\usepackage{fancyhdr}
\usepackage{placeins}
\usepackage{color}
\usepackage{geometry}
\graphicspath{ {Figures/} }
\usepackage{algorithm}
\usepackage[noend]{algpseudocode}
 \usepackage{setspace}

\usepackage[utf8]{inputenc}
\usepackage[english]{babel}

 \usepackage{array}
\usepackage{makecell}

\begin{document}
\baselineskip24pt
\setlength{\thinmuskip}{2\thinmuskip}
\setlength{\medmuskip}{1\medmuskip}
\setlength{\thickmuskip}{1\thickmuskip}
\title{Creating a surrogate commuter network from Australian Bureau of Statistics census data}

\author{Kristopher M. Fair*\textsuperscript{1}, Cameron Zachreson\textsuperscript{1},  Mikhail Prokopenko\textsuperscript{1,2}}

\maketitle
\thispagestyle{fancy}

1. Complex Systems Research Group, School of Civil Engineering, Faculty of Engineering and IT, The University of Sydney, Sydney, NSW 2006, Australia.

2. Marie Bashir Institute for Infectious Diseases and Biosecurity, The University of Sydney, Westmead, NSW 2145, Australia.

{*}corresponding author:
Kristopher M. Fair (kristopher.fair@sydney.edu.au)

\begin{abstract}
Between the 2011 and 2016 national censuses, the Australian Bureau of Statistics changed its anonymity policy compliance system for the distribution of census data. The new method has resulted in dramatic inconsistencies when comparing low-resolution data to aggregated high-resolution data. Hence, aggregated totals do not match true totals, and the mismatch gets worse as the data resolution gets finer. Here, we address several aspects of this inconsistency with respect to the 2016 usual-residence to place-of-work travel data.  We introduce a re-sampling system that rectifies many of the artifacts introduced by the new ABS protocol, ensuring a higher level of consistency across partition sizes. We offer a surrogate high-resolution 2016 commuter dataset that reduces the difference between aggregated and true commuter totals from $\sim34\%$ to only $\sim 7\%$, which is on the order of the discrepancy across partition resolutions in data from earlier years.
\end{abstract}

\section*{Background \& Summary}
High-resolution commuter network information, as well as general information describing population distributions \cite{cite24}, is a major factor in computational modelling of diffusion phenomena in various contexts: demographic \cite{cite19}, epidemiological \cite{cite13, cite14, cite2,cite17}, economic \cite{cite15}, ecological \cite{cite16} and so on. However, privacy constraints on released Census data, in the presence of intricate dependencies between population and employment distributions in relatively small, highly urbanized, but spatially spread countries, such as Australia, coupled with changes in data protocols across census years, present specific challenges in reconstructing commuter (travel-to-work) networks with sufficiently high fidelity \cite{cite21,cite12, cite3, cite4, cite11}.

These challenges manifest in two ways. The first of these pertains to individual microdata, which is organised by household to capture information about both the individual and housing unit. While the collective microdata is a powerful resource, variations in questions asked, possible responses, and record structure often present difficulties in comparing results across years \cite{cite20}. The second challenge relates to the specific methods used by the agencies that gather and report census data, in protecting the anonymity of individuals. While it is necessary for these methods to introduce perturbations, the details of how such perturbations are applied can result in unintended consequences when high-resolution data is aggregated. This is because biases introduced by the perturbation protocol are magnified by aggregation.  

In the recent Australian census datasets \cite{cite18}, these challenges manifest themselves as loss of accuracy in very finely partitioned data, where individual population counts can be on the order 1 to 10 individuals. An important example of such a data set is the commuter network, describing the normal work travel behaviour of the population. The loss of accuracy in such data is primarily due to the specific noise-inducing protocols that the Australian Bureau of Statistics (ABS) employs to ensure the anonymity of census participants. At the same time, this loss in accuracy severely diminishes the usefulness of the commuter networks in modelling contagion phenomena, such as epidemics. In such models, work mobility is a primary driver of contagious diffusion. As such, the accuracy of the commuter network is crucial for realistic outputs regarding aggregate demographic and epidemiological characteristics, such as community and national attack rates. Furthermore, without trustworthy inputs, such models cannot accurately identify salient routes of contagion spread, or analyze mitigation strategies based on network theory.

Similar challenges from noise-inducing protocols, which may also differ across census years, occur in other scenarios in which there is a need to estimate demographic and phenomenological dynamics. This is relevant not only to network-centric studies, but also to more general agent-based simulations, or any study aimed at fine-grained reconstruction of spatio-temporal dynamics \cite{cite22}. Thus, the goal of the present work is not only to reconstruct specific commuter networks of Australia between 2011 and 2016, but also to present a method of microdata reconstruction. The method aims to correct discrepancies that may arise due to Census noise protocols, improving consistency across partition scales while preserving anonymity. The secondary aim is to increase interoperability of Census datasets, in line with the Integrated Public Use Microdata Series (IPUMS) approach \cite{cite20}. 

To further these ambitions we first formalize the network structure and identify discrepancies between different scales of spatial partitioning. We then describe the technical details for constructing our re-sampled network using additional datasets. Finally, we show several comparisons between the ABS provided and re-sampled data that demonstrate the distinction and validity of the resulting dataset.

\begin{figure}[h]
	\centering
	\includegraphics[width=\textwidth]{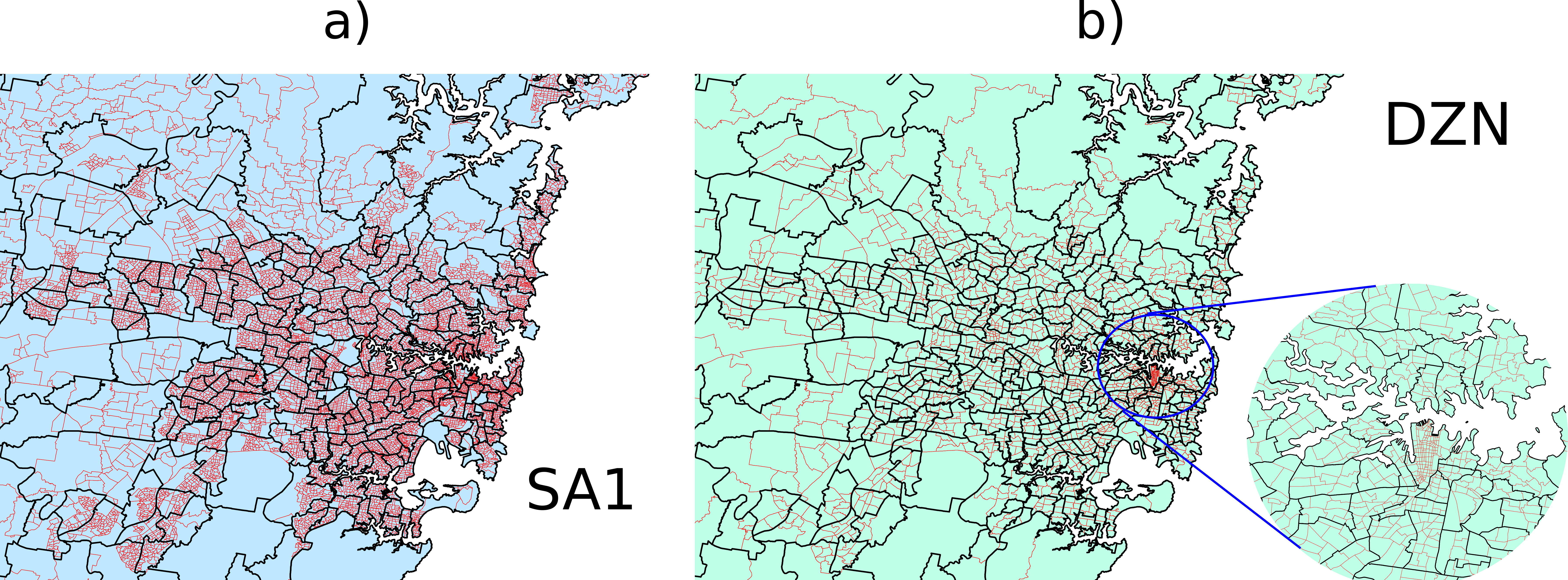}
	\caption{Maps of the Greater Sydney region illustrating the distribution of population partitions. (a) A map of the Greater Sydney region showing SA2 (black) and SA1 (red) population partitions. (b) A map of the same area showing SA2 (black) and DZN (red) partitions. The inset in (b) zooms in on the Sydney central business district to illustrate the much denser packing of DZN partitions in that area. }
	\label{fig_maps}
\end{figure}

The ABS provides access to most census data through the on-line system Census TableBuilder, free of charge, for the 2006 census onward. A subset of the available data is the accumulated microdata of usual-residence (UR) to place-of-work (POW) which constitutes the commuter mobility network (we will refer to this as the TTW, or, travel-to-work dataset). Each census has undergone some re-partitioning of residential and work areas with the latest hierarchical structure divided into four levels of statistical areas for UR ($\text{UR} = [ \text{SA1}, ~\text{SA2}, ~\text{SA3}, ~\text{SA4} ]$), and POW ($\text{POW} = [ \text{DZN}, ~\text{SA2}, ~\text{SA3}, ~\text{SA4} ]$), respectively. This system is defined by the Australian Statistical Geography Standard \cite{cite8}. The smallest of these residential partitions, SA1, is designed to house a population of about 200 to 800 people. Maps of SA2, SA1, and DZN partitions for the Greater Sydney region are displayed in Figure \ref{fig_maps}. SA1 and DZN partitions accumulate to exact partitions on the SA2 scale, this is displayed for SA1 partitions in Fig. \ref{fig_maps}a, and for DZN partitions in Fig. \ref{fig_maps}b. Note that the uneven distribution of employment centres in Australia's cities produces a corresponding non-uniformity in DZN partition density, as displayed in Fig. \ref{fig_maps}b. 

This partitioned commuter data translates to a bipartite network $G_{[\text{UR}\rightarrow \text{POW}]}=(V_G,E_G)$ where $V_G$ is a set of vertices (nodes) of two types $V_G=X\cup Y$, where $X=\{x_1, x_2, ... , x_n\}$ represent the $n$ partitioned UR locations, and $Y=\{y_1, y_2, ..., y_k\}$ represent the $k$ partitioned POW locations. The set of edges 
\begin{equation}
E_G = \{~(x_{i_1}, y_{j_1}),~(x_{i_2},y_{j_2}),~...~(x_{i_{|E_G|}},y_{j_{|E_G|}})~\}\,,
\end{equation}
defines the unique connections between these vertices. For example, UR $x_i$ and POW $y_j$ may be connected by an edge $e_{ij}=(x_{i},y_{j})$. Each subset of edges has a corresponding set of weights, defined by the function: 
\begin{equation}
w_{ij}(\{e_{ij}\}, G)\,, 
\end{equation}
which gives a set of commuter numbers indexed to the corresponding location pairs in $\{e_{ij}\}$, over the network $G$. The use of the argument $G$ is necessary, as the same location pairs may have different numbers of commuters in different networks. For brevity, we will omit the subscripts $i$ and $j$ in cases where they are not required for specificity. We will use similar notation to refer to sets of UR and POW locations associated with edges as $x(\{e\})$ and $y(\{e\})$, as well [Note: the second argument is not necessary here, as the required information is contained in the set $\{e\}$, and does not vary between networks with the same sets of nodes].

As mentioned above, these data sets are subject to a perturbation protocol to prevent cross referencing different variables that may allow the identification of specific individuals \cite{cite7} even with the application of safeguards \cite{cite5,cite6}. Not doing so would violate the Australian Census and Statistics Act 1905 to preserve the anonymity of individuals. This perturbation process is outlined in ABS publications \cite{cite1,cite9, cite10}. 

The sizes of UR and POW population partitions affect the magnitudes of the populations moving between them. Relative to these magnitudes, different levels of noise are required to preserve the anonymity of individuals. For small commuting populations, the perturbation magnitudes will be on the order of the unperturbed values. Furthermore, for the 2016 census, the ABS changed their perturbation protocol by removing a step designed to conserve the total population across different spatial partitions, a property they refer to as `additivity'. Some major practical consequences of removing the additivity-ensuring step are observable discrepancies in the total number of commuters, $N_G =\sum w(E_G, ~G)$, accounted for by the network $G$ on different partition scales. 

\begin{figure}[h]
	\centering
	\includegraphics[width=\textwidth]{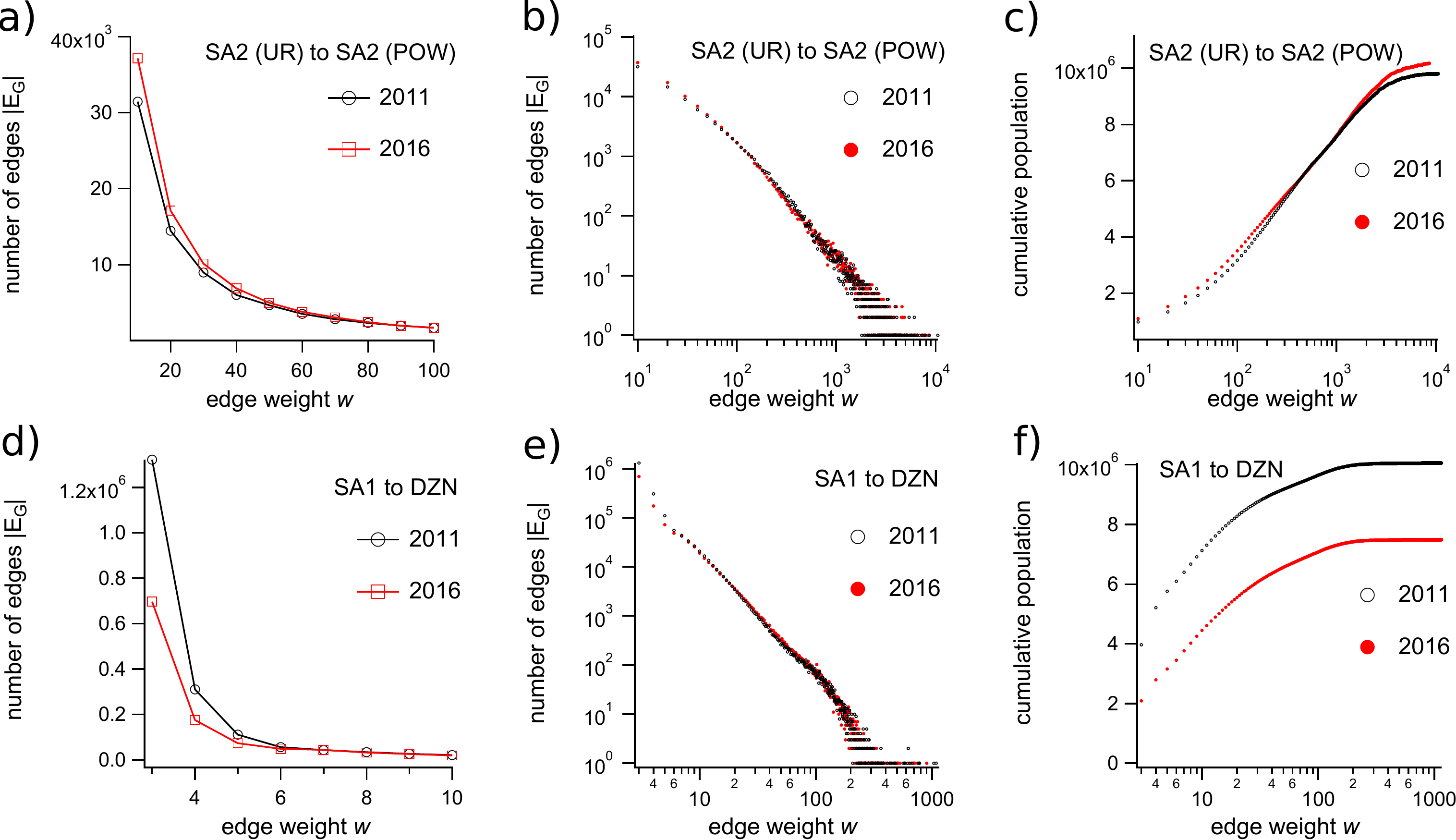}
	\caption{Weight distributions and cumulative population distributions for TTW networks from different census years and partition schemes. (a) Distributions of edge weights ($w < 100$) for the SA2$\rightarrow$SA2 networks for 2011 and 2016, plotted on a linear scale. (b) Distributions of all edge weights for the SA2$\rightarrow$SA2 network from 2011 and 2016 plotted on a log scale. (c) Cumulative population distributions for the SA2$\rightarrow$SA2 network from 2011 and 2016. (d) Distributions of edge weights ($w < 10$) for the SA1$\rightarrow$DZN networks for 2011 and 2016, plotted on a linear scale. (e) Distributions of all edge weights for the SA1$\rightarrow$DZN network from 2011 and 2016 plotted on a log scale. (f) Cumulative population distributions for the SA1$\rightarrow$DZN network from 2011 and 2016. The distributions in (a - c) have bin width of 10, while (c - d) have bin width 1, with a minimum value of 3, artificially introduced by the ABS protocol. The plots in (a) and (d) show only a subset of the weight range, zooming in on the low end of the distribution where the largest discrepancies exist between years.}
\label{fig_WD}
\end{figure}

Edge weight distributions, and cumulative population distributions as a function of edge weight for the SA2$\rightarrow$SA2 and SA1$\rightarrow$DZN commuter networks of 2011 and 2016 are displayed in Figure \ref{fig_WD}. Lower-resolution TTW networks such as those representing connections on the SA2 scale display relatively consistent weight distributions between censuses. Comparison across years shows moderate increases in the numbers of edges across the weight range as could be expected for an increasing employed population between 2011 and 2016 (Fig.~\ref{fig_WD}a and \ref{fig_WD}b). The corresponding distribution of this increased population across the edge weight range is illustrated in Fig.~\ref{fig_WD}c, which does not show any alarming trends or obvious artifacts in the data. Unfortunately, this consistency does not hold for the fine-grained SA1$\rightarrow$DZN network. The weight distributions for this network shown in Fig.~\ref{fig_WD}d and \ref{fig_WD}e indicate a counter-intuitive drop in the numbers of small edges between 2011 and 2016, which corresponds to a dramatic decrease in the total commuting population accounted for by the network. The distribution of the commuting population across the edge weight range (Fig.~\ref{fig_WD}f) confirms that major discrepancies exist between partition schemes, likely due to a significant drop in the number of small edges included in the network. 

As the partitions that comprise the vertices $V_G$ are increasingly subdivided, the weights of the edges connecting them get smaller. The new perturbation protocol appears to dramatically reduce the number of small edges included in the network, particularly around the minimum value of $w=3$. This adversely effects the network both quantitatively, by lowering the commuter populations throughout the network, and structurally, by removing edges from $E_G$, altering the binary structure of the network. In the case of the high-resolution SA1$\rightarrow$DZN network, small edges are a crucial aspect of the network structure, and carry a large portion of the total edge strength. 

\begin{figure}[h]
	\centering
		\includegraphics[width=\textwidth]{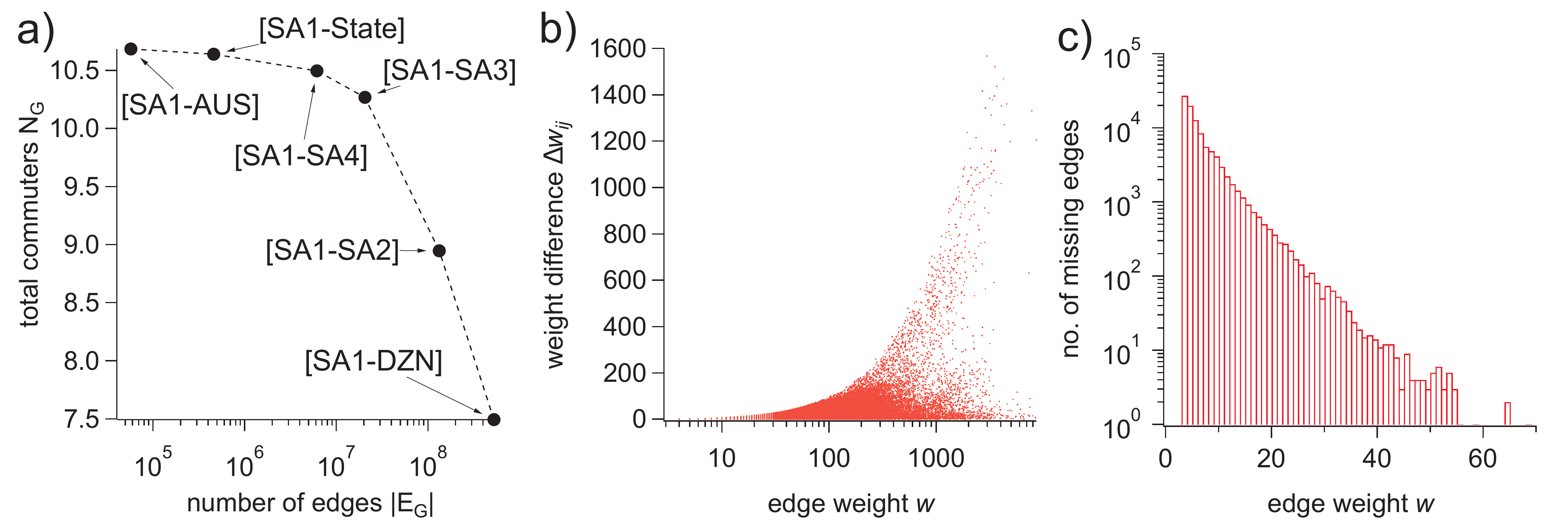}
	\caption{Discrepancies in total population and commuter distribution related to partition aggregation behavior. (a) The total number of commuters $N_G$ in ABS data for networks of varied size. Each point corresponds to a network between SA1 partitions and a different scale of POW partition (national, state, SA4, SA3, SA2, DZN). (b) The discrepancy between commuter numbers, $\Delta w_{ij}$, on each edge $w(E_{AB}, A)$ and $w(E_{AB}, B)$ plotted against $w(E_{AB}, B)$. (c) The frequency distribution as a function of edge weight for edges present in the ABS-provided SA2$\rightarrow$SA2 network ($B$) but not the aggregated SA1$\rightarrow$DZN network ($A$).}
	\label{figure 1}
\end{figure}

The need for a method to ensure consistency in commuter numbers across partition scales is further exemplified in Figure \ref{figure 1}a, which plots the total working population ($N_G$) in networks built by distributing commuters from SA1 partitions into each of the possible POW partition schemes. As the sizes of the POW partitions decrease from the entire nation down to individual destination zones, the total number of commuters drops by $34\%$ while the total number of edges increases by four orders of magnitude.

The structural inconsistency across partition scales that this problem introduces can be understood by amalgamating the vertices of network $G_{[\text{SA1} \rightarrow \text{DZN}]}$ into corresponding SA2 partitions. By doing so, we create network $A_{[\text{SA2} \rightarrow \text{SA2}]}=(V_A, E_A)$, that can be compared to the network constructed from ABS data on the SA2 scale [which we will label network $B_{[\text{SA2} \rightarrow \text{SA2}]}=(V_B,E_B)$]. Network $B$ is missing only $6\%$ of the total commuter population because the edges are composed of more commuters and therefor receive relatively less perturbation from the ABS protocol. This smaller discrepancy is comparable with that of previous years for which the additivity-ensuring step was still included. 

Figure \ref{figure 1}b illustrates the discrepancies between edge weights (commuter numbers between
a given pair of locations) for edges appearing in both networks $A$ and $B$. To compute these discrepancies, we define the set of edges appearing in both $E_A$ and $E_B$ as the intersection $E_{AB} = E_B \cap E_A$, the weights of these edges for networks $B$ and $A$, respectively, as ${\bf w}_B =  w(E_{AB}, ~B)$, and ${\bf w}_A = w(E_{AB}, ~A)$, and the discrepancies $\Delta w$ between the weights of edges existing in both sets
\begin{equation} \label{delta w}
{\Delta w_{ij} = [w_{ij}\in {\bf w}_B] - [w_{ij}\in{\bf w}_{A}]}. 
\end{equation}
 
Using this notation, Fig. \ref{figure 1}b plots $\Delta  w_{ij}$ as a function of $w\in{\bf w}_B$, and demonstrates that the perturbations to small edges in the SA1$\rightarrow$DZN network produce large negative discrepancies in edge weight when the data is aggregated to the SA2$\rightarrow$SA2 scale.

To understand this result in more detail, it is helpful to note that the spatial distribution of the working population is very heterogeneous, with an exponentially larger fraction of the working population employed within central business districts of major cities. However, only the DZN partitions are designed to accommodate this heterogeneity, as they are delineated based on employee population (number of people who work in a region), rather than residential population. On the other hand, SA2 partitions are designed based on residential population which results in a few SA2 business hubs containing many DZN partitions (see Figure \ref{fig_maps}b). In some cases, this results in over $10^3$ component SA1$\rightarrow$DZN edges amalgamating to single, larger SA2$\rightarrow$SA2 edges. 

It is clear that many SA1$\rightarrow$DZN edges are being removed entirely (their weight set to zero) because there are 97,881 edges appearing in the as-provided SA2$\rightarrow$SA2 network $B$ that do not appear after aggregating the SA1$\rightarrow$DZN edges to produce network $A$. This gives $|E_A|\approx0.64|E_B|$ for the SA2-level networks. The frequency distribution for the weights of missing edges, \mbox{$w(\{E_B \setminus E_A\}, B)$} (where the symbol $\setminus$ denotes the set complement), is shown in Figure \ref{figure 1}c which indicates an exponential decrease in removal frequency as a function of edge weight. The data in Figure \ref{fig_WD} and Figure \ref{figure 1} indicate conclusively that many small perturbations on the SA1$\rightarrow$DZN scale accumulate, producing the large discrepancies observed when they are aggregated. 

In this work, we develop and apply a method to restore lost network structure and improve quantitative consistency across commuter networks on different partition scales. The result is a surrogate network $S_{[\text{SA1} \rightarrow \text{DZN}]} = (V_S,E_S)$, on the resolution of SA1 to DZN. This reconstructed commuter network will serve as a platform for ongoing research efforts that utilize Australian travel networks, such as agent-based epidemiological modeling \cite{cite2,Zachresoneaau5294}.

\section*{Methods}
\FloatBarrier

Our method is essentially a re-sampling process that we use to introduce new edges into the SA1$\rightarrow$DZN network to improve quantitative consistency upon aggregation to the SA2 scale. The method does not introduce any new edges to the SA2$\rightarrow$SA2 network upon aggregation, and therefore cannot correct for the missing edges distributed as shown in Fig. \ref{figure 1}c. However, most of the missing commuters are accounted for by correcting the discrepancies shown in Figure \ref{figure 1}b, and our method emphasizes this aspect of the problem. 

Before commencing our procedure, all data provided by the ABS was pre-processed to remove edges that link to non-geographic regions such as "Migratory/ offshore/ shipping" and "No usual address". For the 2016 SA1$\rightarrow$DZN network this accounts for 53,135 edges and 469,854 commuters.  

In addition to the original, perturbed SA1$\rightarrow$DZN network, the method requires the following sets and quantities that we obtained from independent ABS databases:

\begin{itemize}
\item{$N_X = \{N_{x_1}, N_{x_2}, ... N_{x_n}\}$ and $N_Y = \{N_{y_1}, N_{y_2}, ... N_{y_n}\}$, the set of local worker populations for SA1 and DZN partitions, respectively.}
\item{The SA2$\rightarrow$SA2 commuter numbers from the ABS-provided SA2$\rightarrow$SA2 network ($B$).}
\item{The set of (unweighted) SA2$\rightarrow$DZN edges found by creating a mixed-partition network.}
\item{$P(w ~|~ N_x)$, the normalized distribution of edge weights $w$ given residential population $N_x$.}
\end{itemize}

The last item refers to the relationship between local distributions of edge weights and population of the associated SA1, as calculated from 2011 census data obtained without the updated privacy policy compliance protocol.

Our method can be summarized as a two-step process: \par
\begin{enumerate}
	\item Produce a set of $q$ candidate out-edges \\
	$M = \{m_1, m_2, ... m_q\} = \{(x_{i_1}, w_1), (x_{i_2}, w_2)~...~(x_{i_q}, w_q)\}$,
specifying SA1 ($x$) and number of commuters ($w$). This set accounts for the missing workers from each SA1 while maintaining a realistic dependence of weight distribution on UR population $P(w~|~N_x)$.
	\item Build network $S$: add the candidate edges in $M$ into the SA1$\rightarrow$DZN network by specifying DZN ($y$) without violating the topology of the SA2$\rightarrow$DZN network, exceeding the population of the DZN, adding edges that are not present in the SA2$\rightarrow$SA2 network, or exceeding known commuter populations between locations in the SA2$\rightarrow$SA2 network.
\end{enumerate}

In addition to networks $A$, $B$, and $S$ defined above, we will refer to several distinct network sets that are important for the explicit description of our process. For clarity, we will summarize these here and give a brief description of their role in our method:

Network $R$ is the ABS-provided SA1$\rightarrow$DZN network (referred to above as $G_{[SA1\rightarrow DZN]}$), which was released by the ABS subject to the perturbations this work is intended to correct. Network $A$ is the SA2$\rightarrow$SA2 network aggregated from $R$. Network $B$ is the ABS-provided SA2$\rightarrow$SA2 network that exhibits relatively consistent aggregation behavior (that is, the total number of commuters it accounts for is roughly $94\%$ of the known total). We use network $B$ as a quantitative ground-truth while generating the surrogate network. Network $H$ is the ABS-provided SA1$\rightarrow$DZN network from the 2011 census, which exhibits acceptable aggregation behavior. We use network $H$ to build up the set of probability distributions describing $P(w~|~N_x)$. A key assumption of our method is that this relationship between local population and out-edge weight distribution is relatively invariant across census years. Network $\Gamma$ is the ABS-provided SA2$\rightarrow$DZN network which we use as a topological constraint while assigning the candidate edges from each residential zone to appropriate destination zones. That is, we only incorporate SA1$\rightarrow$DZN edges into $S$ that have a corresponding SA2$\rightarrow$DZN pair existing in $\Gamma$. Finally, network $S$ is the surrogate SA1$\rightarrow$DZN network that is the final output of our method and network $C$ is the SA2$\rightarrow$SA2 network aggregated from network $S$.  We compare networks $B$ and $C$ when evaluating the aggregation behaviour of $S$. Some quantitative features of these networks are summarized in Table \ref{Tabel: Networks}. 

\begin{table}[h]
	\caption{Commuter networks and selected characteristics.}
	\vspace{0.25cm}
	\label{Tabel: Networks}
	\centering
	\bgroup
	\def\arraystretch{1.5}
	\begin{tabular}[c]{ |c|c|c|c|c| } 
		\hline
		Network & \makecell{Partition\\ (UR $\rightarrow$ POW)}  & $|E|$ & $\sum w$ & Source \\
		\hline
		$R=(V_R,E_R)$ & SA1 $\rightarrow$ DZN & 1,184,946 & 7,023,571 & ABS 2016\\ 
		\hline
		$A=(V_A,E_A)$ & SA2 $\rightarrow$ SA2 & 118,167 & 7,023,571 & Accumulated from $R$\\ 
		\hline
		$B=(V_B,E_B)$ & SA2 $\rightarrow$ SA2 & 212,805 & 10,073,246 & ABS 2016\\ 
		\hline
		$\Gamma=(V_\Gamma,E_\Gamma)$ & SA2 $\rightarrow$ DZN & 515,250 & 9,853,543 & ABS 2016\\ 
		\hline
		$H=(V_H,E_H)$ & SA1 $\rightarrow$ DZN & 2,046,094 & 10,058,331 & ABS 2011\\ 
		\hline
		$S=(V_S,E_S)$ & SA1 $\rightarrow$ DZN & 1,731,938 & 9,336,333 & Constructed\\ 
		\hline
		$C=(V_C,E_C)$ & SA2 $\rightarrow$ SA2 & 118,167  & 9,336,333 & Accumulated from $S$\\ 
		\hline
	\end{tabular}
	\egroup
\end{table}

The following two sections describe our method in detail. The first describes the process of generating the list of $(\text{SA1}, w)$ pairs which we refer to as ``candidate edges''. The second describes the process of assigning these candidates edges to DZN partitions subject to our selected constraints. 

\FloatBarrier
\subsection*{SA1 candidate edges}\label{sec_candidates} \par 

\begin{figure}[h]
	\centering
	\includegraphics[width=\textwidth]{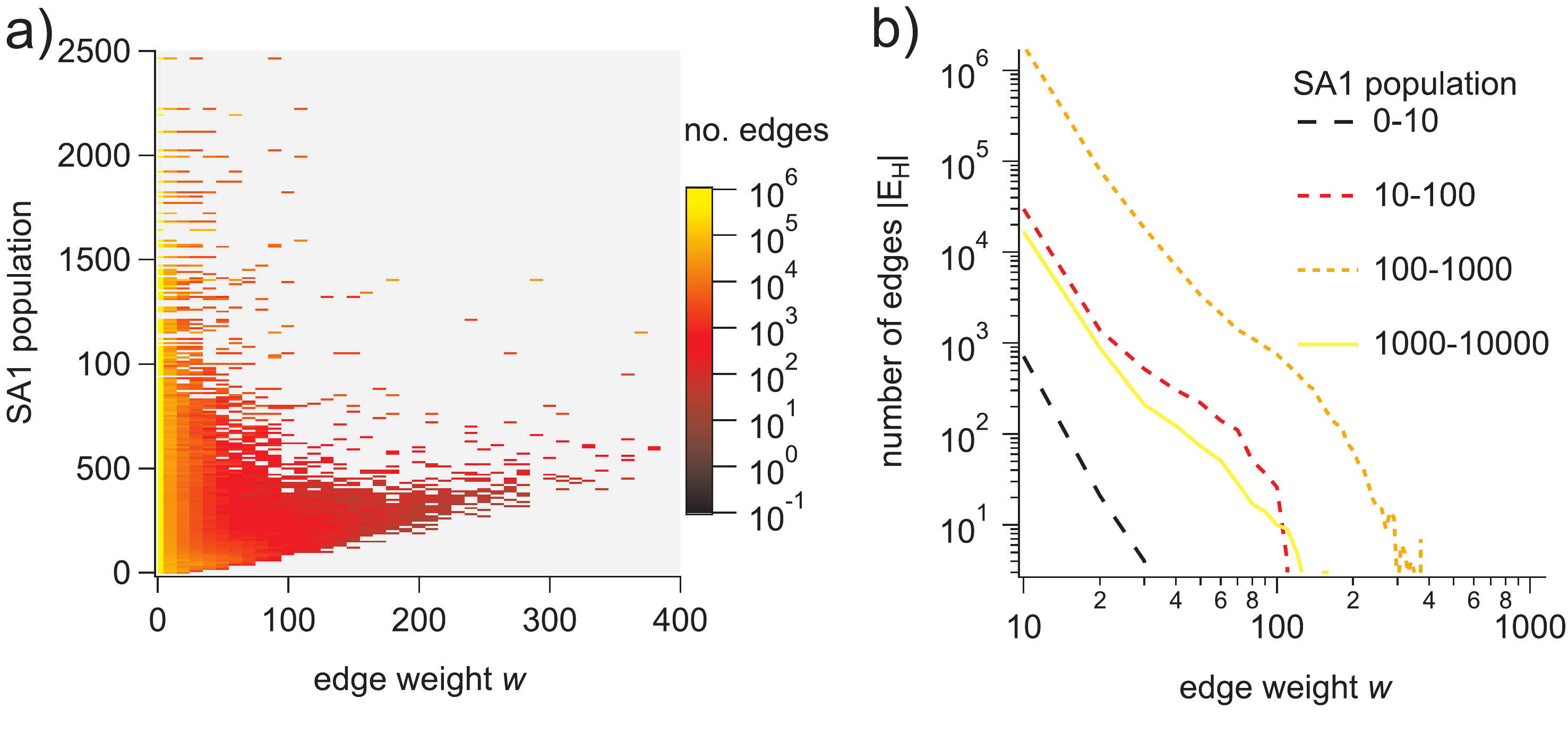}
	\caption{Edge weight frequency distributions as functions of local population. (a) Color plot showing $P(w)$ ($y$ axis) as a function of $N_x$ ($x$ axis) for the 2011 SA1$\rightarrow$DZN commuter network. (b) The frequency distribution of edges as a function of SA1$\rightarrow$DZN commuter network edge weight, where each curve represents the weight frequency distribution for a specific range of SA1 populations.}
	\label{Figure2}
\end{figure}

We observed the behavior of $P(w)$ as a function of $N_x$ to be similar across 2006 and 2011 censuses. This dependence appears to reflect a consistent feature of the commuter mobility network. Although the underlying mechanism producing this set of conditional distributions is not in the scope of this report, it is a subtle aspect of the network structure that should be taken into account. Network $H_{[\text{SA1} \rightarrow \text{DZN}]} = (V_H, E_H)$, derived directly from the 2011 ABS census, along with the 2011 worker populations, gives the distribution of commuter edge weights as a function of the local SA1 population $P(w~|~N_x)$ (shown in Figure \ref{Figure2}). While the method we used to generate these distributions is case-specific, a similar process could be applied in any situation where there is some confidence in the separation of time-scales between real network evolution and artifact introduction due to institutional data processing protocols. Indeed, a more general approach to this aspect of the problem may be needed in cases where true network dynamics are more difficult to distinguish from artifacts. This is an ongoing question that we will continue to address in future work. One promising future direction is to derive a maximum entropy distribution for the weights of the edges leaving each location, constrained by the known numbers of commuters and the worker populations in the destination zones allowed by the topology and SA2$\rightarrow$DZN edge weights of network $\Gamma$. In general, the maximum entropy principle determines the least biased probability distributions, consistent with specific constraints on the average values of measurable quantities \cite{Harding2018}. Other approaches are possible as well, for example, Shannon information could be computed for fragments of the network that exhibit acceptable aggregation behavior, and local weight distributions defined so that sampling from them explicitly addresses information loss in parts of the network adversely affected by the removal of data from the original travel-to-work matrix. Techniques for doing so could be adapted from existing methods where networks are iteratively grown from fragments based on node assortativity constraints, leveraging the relationships between node assortativity and mutual information of the target network \cite{Piraveenan2007,Piraveenan2009}.

Once these conditional distributions are established, we sample from them to account for the number of missing commuters from each SA1. The number of missing commuters associated with a given SA1 partition $x^*$ is computed as the discrepancy between the known working population ($N_{x_i}$) and the sum  $\sum_{j=1}^{k} w(~\{(x^*, y_j)\}, R )$, which is the total out-weight associated with the partition $x^*$. The set of these accumulated populations gives $N_{X_R}$:

\begin{equation}
N_{X_R} = \Biggl\{ ~ \sum_{j=1}^{k} w(~\{(x_1, y_j)\}, R), ~...~ \sum_{j=1}^{k} w(~\{(x_n, y_j)\}, R) \Biggr\}  = \{ ~N_{x_1}^R, ~N_{x_2}^R, ... ~N_{x_n}^R~\}\,,
\end{equation} 
which allows us to calculate the discrepancy in local worker population for each SA1:
\begin{equation}\label{eq_dNx}
\Delta N_X = \{ ~[N_{x_1} - N^R_{x_1}], ~[N_{x_2} - N^R_{x_2}] ,... ~[N_{x_n} - N^R_{x_n}]~\} = \{~\Delta N_{x_1},~ \Delta N_{x_2},...~ \Delta N_{x_n}~\} \,.
\end{equation}

The algorithm then generates $M$ as follows: for each SA1 partition $x_i$, individual weights $w'$ are iteratively sampled from $P(w~|~N_{x_i})$ to produce candidate edges $m' = (x_i, w')$ which are included in $M$ under the condition that 
\begin{equation}
\Delta N_{x_i} > w' + \sum_{m_j\in M} w_j \times \delta(x_{i_j}, x_i) \,,
\end{equation}
where $\delta(x_{i_j}, x_i)$ is equal to 1 if $x_{i_j} = x_i$ and equal to 0 otherwise. If the condition above is not met the candidate edge $m'$ is rejected. The sampling process is repeated until the discrepancies $\Delta N_{X_n}$ are all less than three, the smallest edge size. That is, candidate edges are generated to precisely account for the number of workers missing from each SA1. Quantitative features for an instance of the candidate edge set $M$, and the local populations used to constrain its construction ($N_X$) and assignment ($N_Y$) are shown in Table \ref{Tabel: Sets}. The algorithmic process for creating the set of candidate edges is outlined by the pseudocode in Box \ref{al_candidate}. The following section describes the process of assigning candidate edges to destination zones.

\begin{table}[h]
	\caption{independent data sets and selected characteristics. }
	\vspace{0.25 cm}
	\label{Tabel: Sets}
	\centering
	\bgroup
	\def\arraystretch{1.5}
	\begin{tabular}[c]{ |c|c|c|c|c| } 
		\hline
		Set & Contents & Set size & Total population & Source \\	
		\hline
		\makecell{$M=\{m_1, m_2,... m_q\} =$ \\ $\{(x_{i_1}, w_1), (x_{i_2}, w_2),... (x_{i_q}, w_q)\}$ }& \makecell{SA1 \\candidate\\ edges} & 683,239 & 2,572,117 & Constructed\\ 
		\hline
		$N_X=\{N_{x_1}, N_{x_2},...N_{x_n}\}$ & \makecell{SA1 \\employed\\ residents} & 57,523 & 10,113,273 & ABS 2016\\ 
		\hline
		$N_Y=\{N_{y_1}, N_{y_2},...N_{y_k}\}$ & \makecell{DZN \\employees} & 9,151 & 10,677,111 & ABS 2016\\ 
		\hline
	\end{tabular}
	\egroup
\end{table}

\FloatBarrier
\subsection*{Assigning edges}\label{sec_assigning}

Once the set of candidate edges is generated, each specifying an edge weight and SA1 origin vertex, all that remains is to assign them DZN vertices. Then, the new edges can be included in network $R$ to create the surrogate network $S$. The procedure we used for these assignments is described in this section and outlined in Box \ref{al_assigning}.

We assign candidate edges from $M$ to reasonable DZN partitions by employing $\Gamma_{[\text{SA2} \rightarrow \text{DZN}]}$, $B_{[\text{SA2}\rightarrow \text{SA2}]}$, $E_{AB}$, and $N_Y$ to conditionally restrict the connections that can be added in order to maintain the lower-resolution topology and worker populations at destination zones. The networks $\Gamma$ and $E_{AB}$ are used as binary topological constraints, restricting the possible set of \{SA2, DZN\} and \{SA2, SA2\} location pairs that are compatible with the topology of the new network $E_S$. We use $\Gamma$ as a topological constraint because it represents a good compromise between resolution and quantitative consistency. Because of the larger partitioning of the residential zones $X_\Gamma$, the network loses approximately $8$\% of total commuters due to ABS perturbations, which is much better aggregation behavior than we observe on the SA1$\rightarrow$DZN scale, but worse than the SA2-level network on these terms. On the other hand, it explicitly accounts for connectivity between SA2 residential partitions and DZNs, making it a stronger constraint than the SA2$\rightarrow$SA2 network. We use the overlapping edge set $E_{AB}$ as a topological constraint because it restricts our procedure to those parts of the network in which we have the most confidence. We take this conservative approach in order to avoid introducing edges to the network that could artificially increase connectivity across disparate regions. The local worker populations at each DZN ($N_Y$) are used as quantitative constraints, ensuring that local populations are not exceeded due to the addition of new edges. Similarly, $w(E_{AB}, B)$, the number of commuters between SA2(UR) and SA2(POW) in the portions of network $B$ that overlap with $A$, constrains the number of commuters that can be added to particular edges in $S$.

To select SA1 vertices for the candidate edges $M$, we iterate through the DZN partitions and perform the following procedure: 

For each DZN destination vertex $y_i$ we use $\Gamma$ and $E_{AB}$ to determine the set of possible SA1 origin vertices. These define the subset $M'\subseteq M$ compatible with both the SA2$\rightarrow$DZN and SA2$\rightarrow$SA2 topologies. We then sample $M'$ uniformly at random, combining the sample with the current destination zone $y_i$ to produce a new edge. The new edge is added to the surrogate network under the condition that doing so does not exceed the known number of commuters between SA2 partitions when the surrogate network is aggregated. 

To be precise, $\Gamma$, $E_{AB}$, and $y_i$ define the subset of SA2$\rightarrow$DZN edges 
\begin{equation} \label{eq_E'Gamma}
E'_\Gamma = \{e\in E_\Gamma~|~y(\{e\}) = y_i, (x(\{e\}), \Upsilon_{y_i}) \in E_{AB}\}\,,
\end{equation}
where $\Upsilon_{y_i}$ is the SA2 partition containing the DZN $y_i$. In words, $E'_\Gamma$ is the set of SA2$\rightarrow$DZN edges that point to the destination zone $y_i$ and are consistent with the SA2$\rightarrow$SA2 topology $E_{AB}$. These define the SA2 partitions $\Phi_i  = x( E'_\Gamma)$ and the subset of SA1 partitions contained by them which we will call $X_{\Phi_i}$. From these, the subset of candidate edges is simply determined by selecting only those that contain an element of $X_{\Phi_i}$ as origin vertex:
\begin{equation}
M' = \{m_j \in M~|~x_{i_j} \in X_{\Phi_i}\}\,. 
\end{equation}
Once $M'$ is defined, we randomly select a candidate $m^*\in M' = (x^*, w^*)$ with uniform probability, producing a potential new edge $e^* = (x^*, y_i)$ with weight $w(e^*) = w^*$. The new SA1$\rightarrow$DZN edge $e^*$ aggregates into the SA2$\rightarrow$SA2 edge 
\begin{equation}\label{eq_eB}
e_B = \{e \in E_B ~|~ X_x \supseteq x(\{e^*\}), ~Y_y \supseteq y_i\} = (x_B, y_B) \,,
\end{equation}
where $X_x$ and $Y_y$ are the sets of SA1 and DZN zones contained (respectively) by the SA2(UR) and SA2(POW) partitions in each element of $E_B$. 

To check whether or not the new edge $e^*$ should be added to the surrogate network, we aggregate $E_S$ over the SA1 and DZN vertices contained by the SA2 partitions $x_B$ and $y_B$, and determine whether adding the new edge will exceed the known number of commuters between SA2 zones. That is, the edge $e^*$ is added to $E_S$ under the condition that 
\begin{equation}
w(\{e_B\}) \geq w(\{e^*\}) + \sum_{e_{ij}\in E_S} w(\{e_{ij}\}, S) \times \delta (e_{ij}, X_{x_B}, Y_{y_B}) \,,
\end{equation}
where $X_{x_B}$ and $Y_{y_B}$ are the sets of SA1 and DZN partitions contained by the SA2(UR) and SA2(POW) zones specified by $x_B$ and $y_B$, respectively, and 

\begin{equation}
\delta (e_{ij}, X_{x_B}, Y_{y_B}) = 
\begin{cases}
~1~, & ~\text{if} ~ x_i \in X_{x_B}~\text{AND}~~y_j \in Y_{y_B}\\
\\
~0~, & \text{otherwise}
\end{cases}
\end{equation}
\vspace{0.1cm}

\begin{figure}[h]
	\centering
	\includegraphics[width=0.5\textwidth]{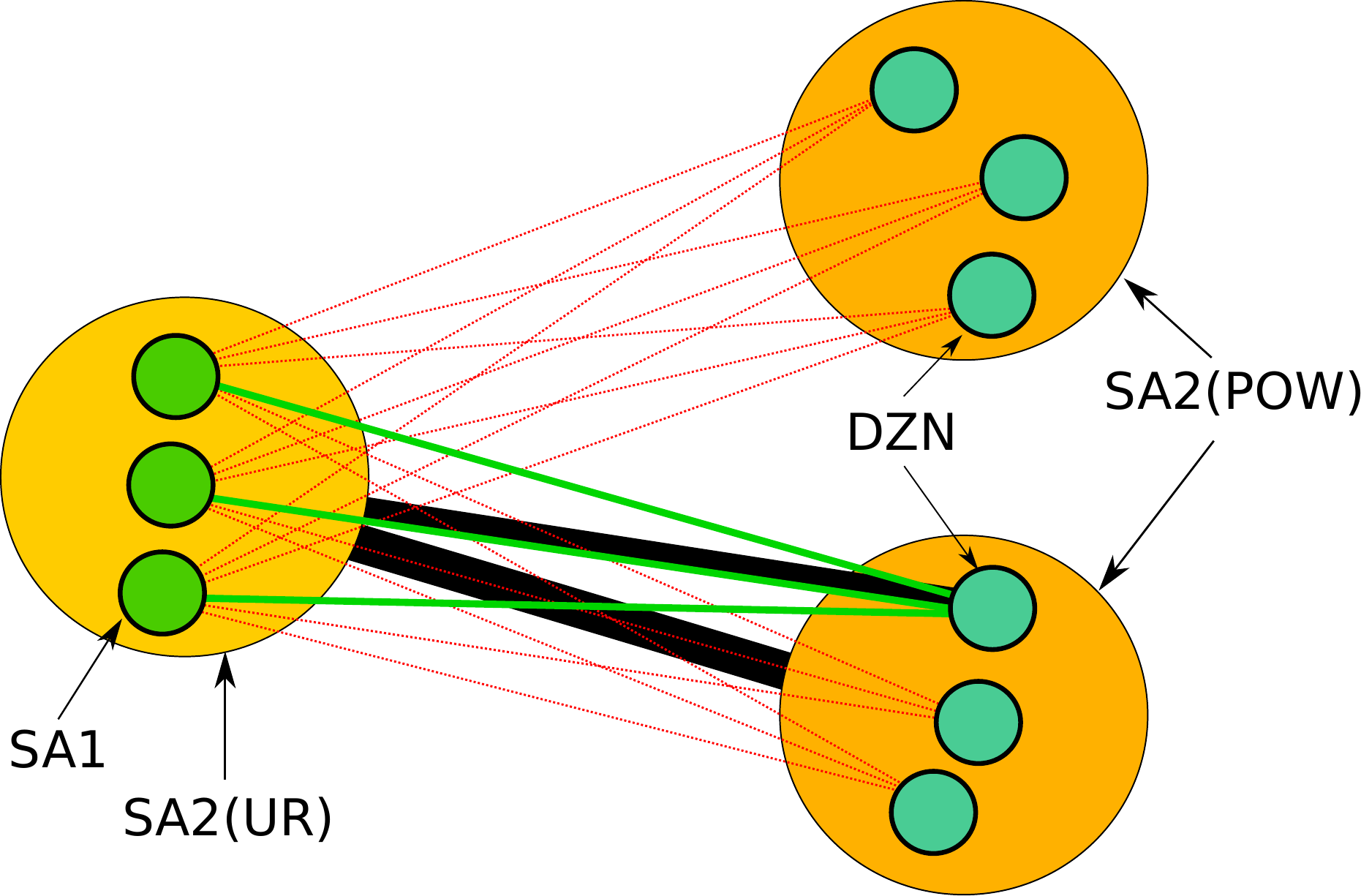}
	\caption{Schematic of topological constraints applied when adding new edges to the surrogate network. The black lines represent the known SA2$\rightarrow$SA2 and SA2$\rightarrow$DZN connections given by networks $B$ and $\Gamma$. The green lines are allowed surrogate SA1$\rightarrow$DZN edges, as they are consistent with the known larger-scale topology. The red lines represent edges that are not allowed, as their inclusion would violate our constraints after aggregation of the surrogate to larger partition schemes.}
	\label{topo_constraints}
\end{figure}

To summarize, the algorithm allows addition of $e^*$ to $E_S$ if aggregation of $E_S$ to larger partitions only produces edges that already exist in $E_\Gamma$ and $E_{AB}$, these topological constraints are illustrated in figure \ref{topo_constraints}. Aggregated edge weights are constrained as well, so that addition of $w(\{e^*\})$ does exceed the value given by $w(\{e_B\}, B)$ upon aggregation of $E_S$ to the SA2$\rightarrow$SA2 scale. After successful assignment of edge $e^*$ into $E_S$, the candidate edge $m^*$ is removed from $M$ and the process is repeated until edges meeting this condition cannot be found.

In principle, the above criterion is sufficient to ensure self-consistency across differently-partitioned data sets, however, the criteria must still account for the effect of the privacy policy compliance perturbations. To account for possible mismatch between employee numbers, we added the additional criterion that the number of workers assigned to destination $y_i$ must not exceed local worker population $N_{y_i} \in N_Y$. Therefore, the condition
\begin{equation}
N_{y_i} \geq w(\{e^*\}) +  \sum_{e_{ij}\in E_S} w(\{e_{ij}\}, S) \times \delta (y(\{e_{ij}\}), y_i) \,,
\end{equation}
must be met, or the edge is not added to $E_S$. Here, $\delta (y(\{e_{ij}\}), y_i)$ is equal to 1 if $y(\{e_{ij}\}) = y_i$, and equals 0 otherwise.

\FloatBarrier
\begin{algorithm}
\floatname{algorithm}{Box}
	\caption{: Candidate edge set algorithm. Pseudocode for the algorithm that produces a list of candidate edges from each SA1 that match the local commuter populations and dependence of edge weight distribution on worker population as determined by the 2011 census.}\label{al_candidate}
	\begin{algorithmic}\setstretch{1.5}
		\Procedure{Generate candidate edges}{}
		\State \textbf{input:}
		\State \indent $N_{X_R}$, the number of SA1 employees aggregated from $R$
		\State \indent $N_X$, the number of SA1 employees reported by ABS 
		\State \indent $P(w|N_x)$, the 2011 edge weight distribution conditional on local population 
		\State \textbf{for} $x_i$ in $X_R$:
		\State \indent $N^R_{x_i} =\sum_{m}^{j=1} w(\{(w_i,y_j)\},R)$
		\State \indent $\Delta N_{x_i} = N_{x_i} - N^R_{x_i}$, the number of employees remaining unassigned from $x_i$ 
		\State \indent \textbf{while} $\Delta N_{x_i}~>~3$ \textbf{do:}
		\State \indent \indent $w'~=~sample~w~with~probability~P(w|N_{x_i})$	
		\State \indent \indent \textbf{if:} $\Delta N_{x_i}\ge w'$
		\State \indent \indent \indent $m' = (x_i,w')$
		\State \indent \indent \indent append $m'$ to $M$
		\State \indent \indent \indent $\Delta N_{x_i}~~ -$=$ ~~w'$, subtract $w'$ from $\Delta N_{x_{i}}$
		\State \indent \indent  \textbf{end if}
		\State \indent  \textbf{end while}
		\State  \textbf{end for}
		\EndProcedure
	\end{algorithmic}
\end{algorithm}
\begin{algorithm}
\floatname{algorithm}{Box}
	\caption{: Destination assignment algorithm. Pseudocode for the algorithm that links the candidate edges to DZN partitions, producing the surrogate network $S$.}\label{al_assigning}
	\begin{algorithmic}\setstretch{1.5}
		\Procedure{Assigning candidate edges}{}
		\State \textbf{input:}
		\State \indent $E_B$, the SA2(UR)$\rightarrow$SA2(POW) network reported by ABS 
		\State \indent $\Gamma$, the SA2$\rightarrow$DZN network
		\State \indent $M$, the candidate edges produced by Algorithm 1
		\State \indent $R$, the SA1$\rightarrow$DZN network reported by ABS
		\State \indent $N_Y=\{N_{y_1}, N_{y_2},...N_{y_k}\}$, the DZN employee population
		\State
		\State initialize $S = R$
		\State initialize $\{\Delta w\}$, the discrepancies in aggregated commuter numbers (see equation \ref{delta w})
		\State \textbf{while} $|M| > 1$
		\State \indent \textbf{for} $y_i$ in $Y_R$:
		\State \indent \indent $E'_\Gamma = \{e\in E_\Gamma~|~y(\{e\}) = y_i, (x(\{e\}), \Upsilon_{y_i}) \in E_{AB}\}$ (equation \ref{eq_E'Gamma})
		\State \indent \indent $\Phi_i = x(E'_\Gamma)$, the SA1 partitions contained by the SA2(UR) partitions of $E'_\Gamma$	
		\State \indent \indent $M' = \{m_j\in M~|~x_{i_j}\in X_{\Phi_i}\}$, subset of $M$ such that $\Phi_i$ contains $x_{i_j}$
		\State \indent \indent $sample~m^* = (x^*, w^*)~from~M'~uniformly~at~random$
		\State \indent \indent $e^* = (x^*,y_i)$,  $w(\{e^*\}) = w^*$,  the potential new SA1$\rightarrow$DZN edge
		\State \indent \indent $e_B = \{e \in E_B ~|~ X_x \supseteq x(\{e^*\}), ~Y_y \supseteq y_i\} = (x_B, y_B)$ (equation \ref{eq_eB})
		\State \indent \indent \textbf{if:} $w(\{e^*\}) > \Delta w(\{e_B\})$ AND $N_{y_i} \ge w(\{e^*\})+\sum_{p=1}^n w(\{(x_p,y_i)\},S)$
		\State \indent \indent \indent append $e^*$ to $E_S$ 
		\State \indent \indent \indent $\Delta w(\{e_B\}) ~~-$=$~~ w(\{e^*\})$
		\State \indent \indent  \textbf{end if}
		\State \indent \textbf{end for}
		\State \textbf {end while}		
		\EndProcedure
	\end{algorithmic}	
\end{algorithm}

\FloatBarrier
Of the 2,572,117 commuters accounted for by the full set of 683,239 candidate edges $M$, there were 729,209 commuters comprising 61,855 edges remaining unassigned when our process terminated due to an inability to assign edges under the above criteria. Two factors are responsible for the inability of the algorithm to assign these edges. The first is that the privacy protocol, by design, ensures cross referencing totals do not match in perturbed data released by the ABS. The second is that our ground-truth topology omits the non-overlapping set $w(\{E_B \setminus E_{A}\}, B)$, therefore, the 612,215 missing commuters tabulated in Figure \ref{figure 1}c cannot be accounted for by our re-sampling procedure.

This surrogate network has an additional 546,992 SA1$\rightarrow$DZN edges, a $25\%$ increase as compared to network $R$, with a total number of commuters $N(S)$ comparable to that of the SA2$\rightarrow$SA2 network, $N(B)$. The total number of commuters in the as-provided SA1$\rightarrow$DZN network $N(G)$ is 7,023,571 the total for the surrogate network $N(S)$ is 9,336,333 and our quantitative ground-truth $N(B)$ is 10,073,246.

\FloatBarrier
\subsection*{Code availability}
The custom code used to generate the surrogate network \textit{via} the method outlined in this text was run on MATLAB version R2017b. The script and required inputs can be accessed on the online repository \cite{cite23}, along with usage notes and descriptions of relevant parameters.

\FloatBarrier
\section*{Data Records}
We have made an instance of the reconstructed surrogate commuter network publicly available \cite{cite23}. All of the data sets used including the original SA1$\rightarrow$DZN commuter mobility network, SA2$\rightarrow$DZN network, SA2$\rightarrow$SA2 mobility network, number of employees in each SA1 ($N_X$), number of employees in each DZN ($N_Y$), SA1 to SA2 correspondence files, and DZN to SA2 correspondence files are publicly available for both 2011 and 2016 through either Census TableBuilder (http://www.abs.gov.au/websitedbs/D3310114.nsf/Home/2016\%20TableBuilder) or the ABS website (http://www.abs.gov.au/). The 2011 SA1$\rightarrow$DZN network ($H$) is no longer publicly available with the additivity-including privacy policy compliance protocol so we provide the version we used along with our surrogate network. The stability of the files available through ABS may vary with time, as evident in the removal of the additivity-ensuring step from the perturbation protocol used for all presently distributed data. 

\section*{Technical Validation}

\begin{figure}[h]
	\centering
	\includegraphics[width=\textwidth]{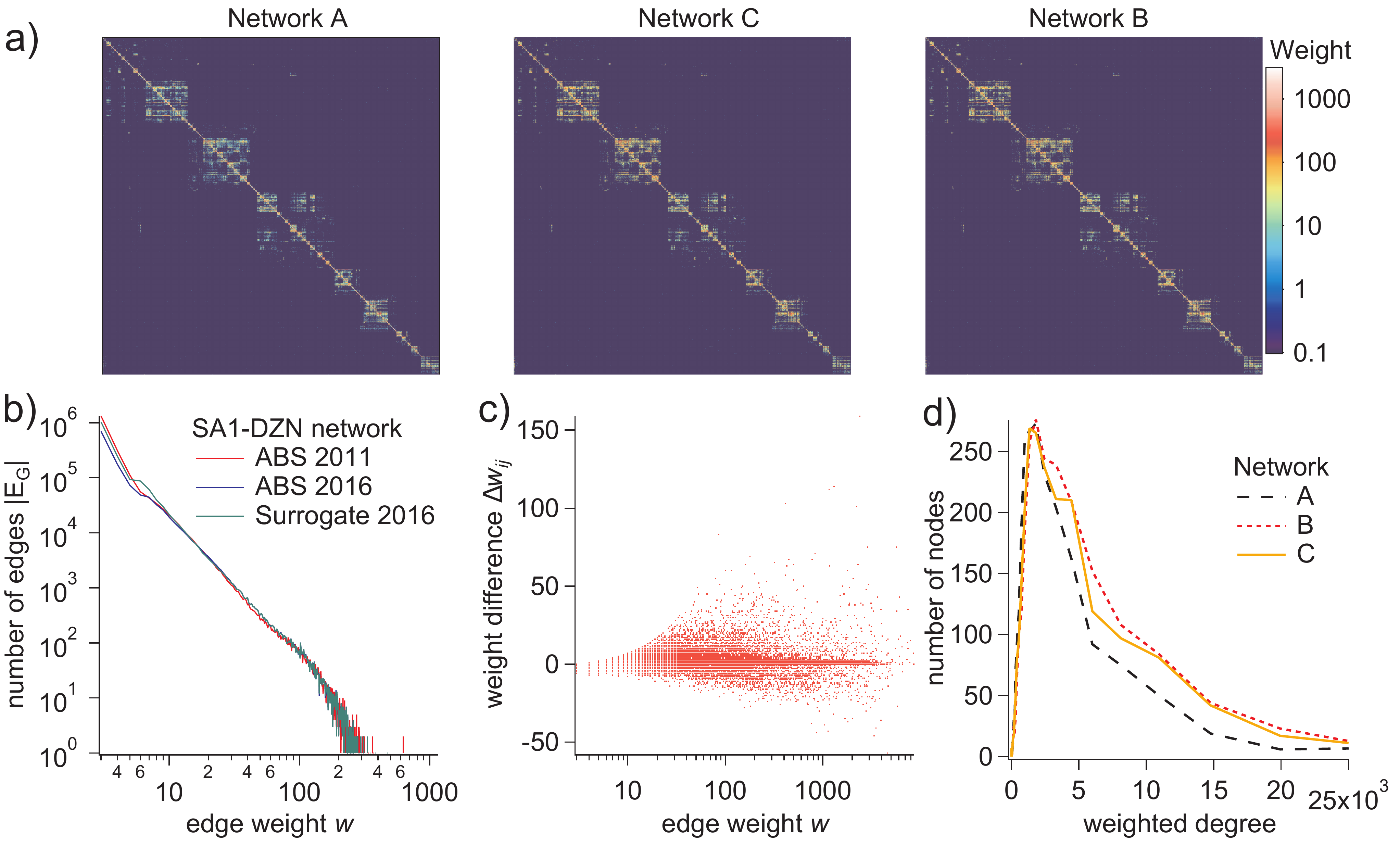}
	\caption{Validation of the surrogate network. (a) Color plots of the SA2$\rightarrow$SA2 adjacency matrix from the aggregate of the original SA1$\rightarrow$DZN network $A$, aggregated surrogate $C$, and ABS-provided SA2$\rightarrow$SA2 network $B$. The SA2 regions are somewhat spatially ordered such that the different states, in particular the larger urban areas, are clustered around the diagonal. (b) Weight distributions for the networks $R$, $H$ and $S$. (c) Weight differences, $\Delta w_{ij}$, as a function of $w(E_{AB}, B)$, demonstrate improved quantitative agreement (compare to Fig. \ref{figure 1}b). (d) Distributions of node degree strength (total incident edge weight) for networks $A$, $B$, and $C$. }
	\label{figure 4}
\end{figure}

To quantitatively assess the aggregation behavior of the surrogate network $S$, we first accumulated its component edges into the corresponding SA2$\rightarrow$SA2 topology (which we will refer to as network $C$). This new aggregated surrogate network was then compared to both the ABS-provided SA2$\rightarrow$SA2 network and the aggregate of the original SA1$\rightarrow$DZN network ($A$), by several different metrics. To assess the overall agreement between the three networks, we first translated their edge lists and weights into adjacency matrices (Figure~\ref{figure 4}a), and computed the 2D correlation coefficient between each pair:
\begin{equation}\label{eq_2Dcorr}
r(\alpha, \beta) = \frac{\Sigma_m\Sigma_n(\alpha_{mn}-\bar{\alpha})(\beta_{mn}-\bar{\beta})} {\sqrt{\strut{3}\Sigma_m\Sigma_n(\alpha_{mn}-\bar{\alpha})^2 \Sigma_m\Sigma_n(\beta_{mn}-\bar{\beta})^2 }},
\end{equation}
where $\alpha$ and $\beta$ represent each of the two adjacency matrices being compared. This comparison demonstrates a high degree of similarity between all three networks, with a significant improvement in correlation between the ABS-provided SA2$\rightarrow$SA2 network and the accumulated surrogate (Table~\ref{Tabel: corr2}).

\begin{table}[h]
	\caption{2D correlation coefficients computed according to Eq. \ref{eq_2Dcorr}, between aggregated and ABS-provided SA2$\rightarrow$SA2 networks. Network $C$ is the aggregated SA1$\rightarrow$DZN surrogate, network $A$ is the aggregate of the SA1$\rightarrow$DZN network provided by ABS, and network $B$ is the SA2$\rightarrow$SA2 network provided by ABS.  \ }
	\vspace{0.25 cm}
	\label{Tabel: corr2}
	\def\arraystretch{1.5}
	\centering
	\begin{tabular}[c]{ |c|c|c|c| } 
		\hline
		 Network pair & $B, A$ & $B, C$ & $ A, C$  \\
		\hline
		2D correlation ($r$) & 0.9821 & 0.9996 & 0.9828 \\  
		\hline
	\end{tabular}
\end{table}

Plotting the frequency distribution of edge weights for the ABS-provided SA1$\rightarrow$DZN commuter networks of 2016 and 2011, along with the corresponding distribution for the surrogate network (Figure \ref{figure 4}b) indicates a partial repair of the discrepancy in low-weight ($w < 10$) edge numbers observed between 2011 and 2016 (Fig.~\ref{fig_WD}d). 

The discrepancies in edge weights between the amalgamated surrogate network ($C$) and the ABS-provided SA2$\rightarrow$SA2 network ($B$) are plotted in Figure~\ref{figure 4}c as a function of the edge weight from network $B$. Comparison of these discrepancies to those plotted in Figure \ref{figure 1}b indicates a dramatic improvement, comparable to the corresponding discrepancies computed for the 2011 commuter network. To further demonstrate the structural repair imparted to the surrogate network, we computed the distributions of weighted degree (the sum of all edge weights incident on each node), for networks $A$, $B$, and $C$ (Figure \ref{figure 4}d). The distribution corresponding to the aggregated surrogate network more closely matches that of the raw SA2$\rightarrow$SA2 network.

We further quantify the similarity between our amalgamated surrogate ($C$) and the ground-truth network (the edges in network $B$ that also exist in network $A$), by calculating the mean-squared error (MSE) in the weights over all UR$\rightarrow$POW pairs in $E_{AB}$. Here, we compute the MSE over the edge weight sets 
\begin{equation}
\alpha = w(E_{AB}, B)\,,
\end{equation}
and 
\begin{equation}
\beta = w(E_{AB}, C) ~~{\text{or,}}~~ \beta = w(E_{AB}, A)\,,
\end{equation}
as
\begin{equation}\label{eq_mse}
\text{MSE}(\alpha, \beta) =\frac{1}{|E_{AB}|}\sum\limits_{e_{ij}\in E_{AB}} [\alpha_{ij} - \beta_{ij})]^2 \,,
\end{equation}
where subscripts $ij$ indicate specific UR$\rightarrow$POW pairs. This quantity provides an estimate of how much our algorithm rectified discrepancies between SA2$\rightarrow$SA2 edges, given our conservative choice not to add edges to the overlapping set $E_{AB}$. The results are shown in Table~\ref{Tabel: MSE} below, and indicate a significant quantitative improvement, as expected from comparison between Figure~\ref{figure 1}b~and~Figure~\ref{figure 4}c. 

\begin{table}[h]
	\caption{MSE between overlapping portions of the aggregated and ABS-provided SA2$\rightarrow$SA2 networks computed according to equation \ref{eq_mse}. \ }
	\vspace{0.25 cm}
	\label{Tabel: MSE}
	\def\arraystretch{1.5}
	\centering
	\begin{tabular}[c]{ |c|c|c|c| } 
		\hline
		 Network pair & $B, A$ & $B, C$ & $ A, C$  \\
		\hline
		MSE & 62.51 & 0.27 & 60.93 \\  
		\hline
	\end{tabular}
\end{table}
To evaluate the improvement in structural properties of the surrogate relative to the as-provided network we analysed two key network measures for the common components of the networks $A$ and $B$. The first is simply the average shortest path between nodes, as computed by applying Dijkstra's shortest-path algorithm to the weighted networks, interpreting edge weight as inverse distance. The second is a version of the clustering coefficient adapted to weighted networks \cite{PhysRevE.71.065103} that defines the weighted clustering coefficient for a node $i$ by evaluating the fraction of its neighbors $j$ and $k$ that share connections, weighted based on the relative weights of the edges connecting the triangle, as
\begin{equation}\label{eq_clustering}
C_i = \frac{2}{k_i(k_i-1)}\sum\limits_{j,k}(\hat{w}_{ij}\hat{w}_{jk}\hat{w}_{ki})^{1/3} \,,
\end{equation}
and reports the average of this quantity over all nodes in the network. Here, the weights of nodes in a triangular cluster are scaled by the largest weight in the network $\hat{w}_{ij}=w(\{e_{ij}\})/max(w(E))$, and $k_v$ is the degree of node $v$. 

\begin{table}[h]
	\caption{Average weighted network statistics. The networks marked with an asterisk ($^*$) contain only edges appearing in $E_{AB}$ that is, they represent the overlapping portions of the networks. [Note: inclusion of the edges unique to network $B$ quantified in Fig. \ref{figure 1}c, produces a dramatic reduction in the network's clustering coefficient, which is intuitive given the relatively low weights of these edges and our definition of the weighted clustering coefficient (Eq. \ref{eq_clustering}).] }
	\label{Tabel: network statistics}
	\vspace{0.25cm}
	\centering
	\begin{tabular}[c]{ |c|c|c|c|c| } 

		\hline
		 Network & $A^*$ & $C^*$ & $B^*$ & $B$  \\
		\hline
		Shortest path & 0.157  & 0.118 & 0.099 & 0.095 \\ 
		\hline
		Clustering coefficient ($\times10^-3$)& 1.97 & 2.95 & 3.11 & 1.51\\  
		\hline
	\end{tabular}
\end{table}

These network statistics are shown in Table \ref{Tabel: network statistics} and indicate improved correspondence between the network properties of the overlapping sets $w(E_{AB}, C)$ and $w(E_{AB}, B)$, as compared to the aggregate of the original network $w(E_{AB}, A)$.

The number of commuters in the surrogate network is 9,336,333 constituting a 25\% increase in the commuter population as compared to the aggregated ABS-provided SA1$\rightarrow$DZN network. Our procedure added nearly half a million new SA1 to DZN edges. The increase in correlation and closer network statistics at the SA2 scale, as well as the edge-wise decrease in mean-squared error indicates both a quantitative and structural improvement over the original dataset provided by the ABS. 

The surrogate network proffered here represents a significant improvement over the original SA1 partitioned commuter mobility network. It reconstructs the population and network statistics of the less perturbed SA2 level network by adding additional  SA1$\rightarrow$DZN connections that have been lost to the ABS privacy protocol. Access to the surrogate network and the availability of a method for recovering high fidelity data on such high resolution networks is of broad significance to the computational modeling of diffusion phenomena in various disciplines.
The redistribution of ABS data is protected under Creative Commons licensing.

\subsection*{Network statistics for different instantiations}
The process of generating the surrogate networks is stochastic. However, the constraints placed on the new edge generation leads to very consistent surrogate network statistics across instantiations. This is evident in comparing the network statistics of the surrogate network analysed here, with several additional instantiations. These are shown in Table \ref{Tabel: network statistics for additional surrogates}.   
\begin{table}[h]
	\caption{The weighted network statistics for additional surrogate data sets.}
	\label{Tabel: network statistics for additional surrogates}
	\vspace{0.25cm}
	\centering
	\begin{tabular}[c]{ |c|c|c|c|c| } 
		
		\hline
		Network & $C$ & $C1$ & $C2$ & $C3$  \\
		\hline
		Shortest path & 0.119  & 0.113 & 0.116 & 0.119 \\ 
		\hline
		Clustering coefficient & 2.70 & 2.70 & 2.65 & 2.70 \\  
		\hline
	\end{tabular}
\end{table}

Likewise the MSE and 2D correlation demonstrate an excellent agreement between the specific surrogate network analysed produced by our study, and additional generated surrogates. These are shown in Table \ref{Tabel: numeric compatison for additional surrogates}. 
\begin{table}[h]
	\caption{The MSE and 2D correlation between the chosen surrogate, C, and additional generate surrogate networks aggregated to SA2$\rightarrow$SA2.}
	\label{Tabel: numeric compatison for additional surrogates}
	\vspace{0.25cm}
	\centering
	\begin{tabular}[c]{ |c|c|c|c| } 
		
		\hline
		Network & $C1$ & $C2$ & $C3$  \\
		\hline
		MSE & 0.142 & 0.153 & 0.148 \\ 
		\hline
		2D correlation & 1.000 & 1.000 & 1.000 \\  
		\hline
	\end{tabular}
\end{table}

\subsection*{Convergence}

\begin{figure}[h]
 	\centering
 	\includegraphics[width=0.49\textwidth]{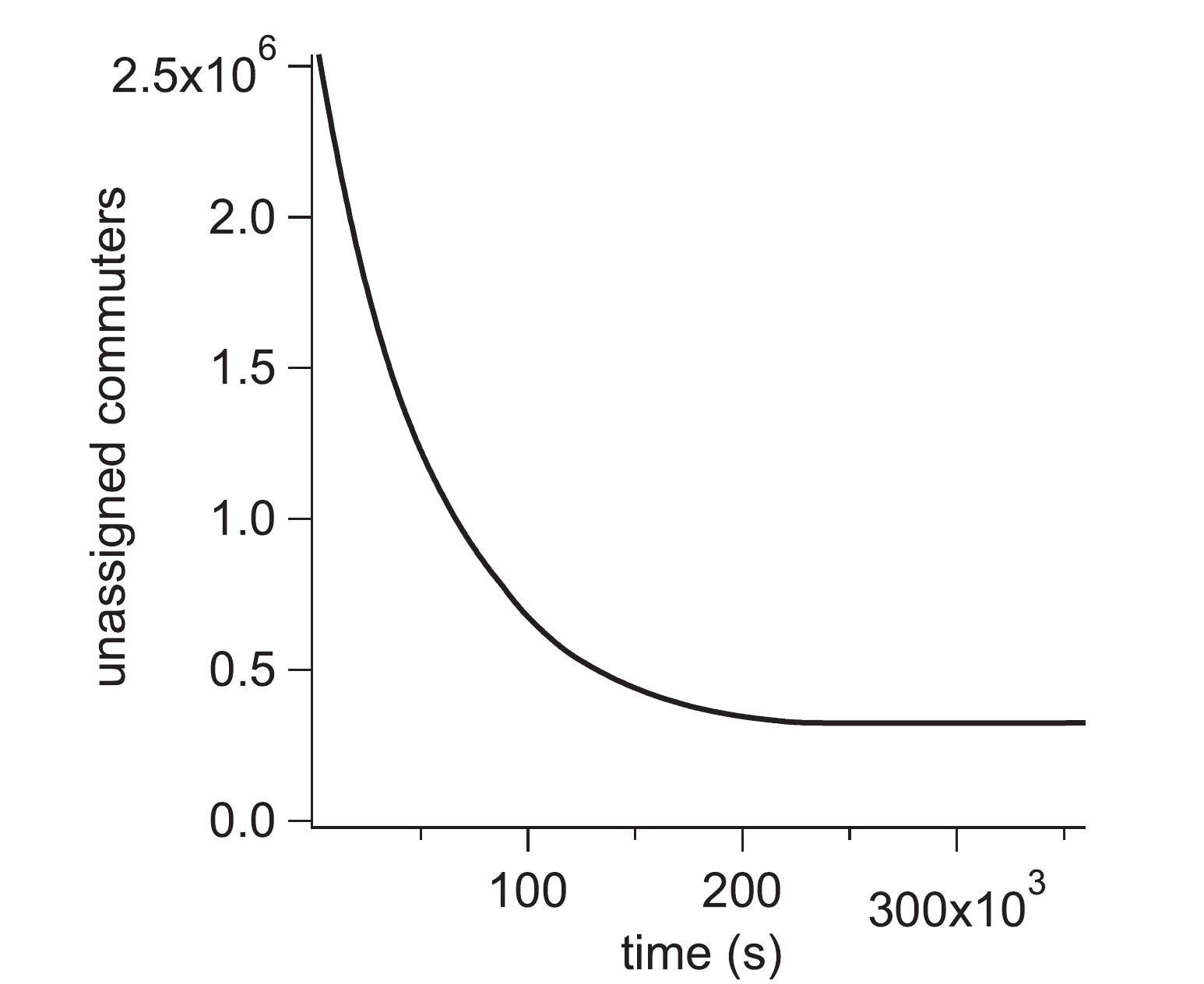}
 	\caption{Algorithm convergence. The number of unassigned commuters as a function of time while assigning commuter weights to new SA1-DZN edges, running the script `create\_surrogate.m' \cite{cite23} for 100 hours.}
 	\label{Appendix_time_convergence}
 \end{figure}

The process of building the new edges $e^*$ from the sample edge distributions is the most time consuming part of creating the surrogate networks. Each run generating a surrogate network was given 100 hours to reach the end-point criteria, however a small proportion of commuters remain impossible to assign, as larger candidate edges become disallowed by the algorithm's constraints. Figure \ref{Appendix_time_convergence} shows the number of unassigned commuters as a function of time when placing the new edges. As edges are added the constraints of SA1 population, DZN population and SA2-SA2 edge sizes reduce the likelihood of a suitable sample edge fitting. This results in convergence on a non-zero number of unassigned commuters.

\FloatBarrier
\section*{Usage Notes}
The MATLAB script `creating\_surrogate.m', available in the online repository \cite{cite23} implements the method outlined in this paper. The input required for this script is located in the repository file `inputs.mat'. This workspace includes:
\begin{itemize}
	\item 2016 SA1-DZN commuter network ($R$),
	\item 2011 SA1-DZN commuter network ($H$),
	\item 2011 SA1 UR populations,
	\item 2016 SA1 employed residents ($N_X$),
	\item 2016 DZN employees ($N_Y$),
	\item 2016 SA2-DZN ABS network ($\Gamma$),
	\item SA2-SA2 network accumulated from $R$ ($A$),
	\item SA2-SA2 ABS network ($B$).
\end{itemize}	
Using this script first produces the commuter residential distribution based on the 2011 census data, then a list of possible SA1 edges ($M$) using the residential distribution and finally assigns them to DZNs, creating $e^*$. This is then combined with the existing edges of network $R$ to create the surrogate network $S$. A complete description of each network and the file header information is located in the corresponding 'README.txt'. The data format is simply a table of edges, the first column corresponding to the SA1 code, the second column corresponding to the DZN code, and the third column giving the number of commuters assigned to the pair. 

\section*{Acknowledgments} 
We acknowledge the Australian Bureau of Statistic (ABS) for providing all of the raw data as well as general advice in regards to the nature of their perturbation procedures. The Authors were supported through the Australian Research Council Discovery Project DP160102742.

\section*{Author Contributions}
KF, CZ, and MP designed the research; KF and CZ designed the algorithm; KF implemented the algorithm code; CZ and KF designed the validation strategy; KF performed data analysis for validation; CZ, KF, and MP composed the manuscript. 

\section*{Competing Interests}
The authors declare no competing interests.

\FloatBarrier

\end{document}